\documentclass[english,showpacs,superscriptaddress,prl,preprintnumbers,amsmath,amssymb,floatfix,reprint]{revtex4-1}
\usepackage[T1]{fontenc}
\usepackage[utf8]{inputenc}
\usepackage{color}
\usepackage{babel}
\usepackage{amsmath}
\usepackage{graphicx}
\usepackage{esint}
\usepackage[unicode=true,pdfusetitle,
 bookmarks=true,bookmarksnumbered=false,bookmarksopen=false,
 breaklinks=false,pdfborder={0 0 0},pdfborderstyle={},backref=false,colorlinks=true]
 {hyperref}

\makeatletter
\usepackage{dcolumn}
\usepackage{bm}
\usepackage{color}
\usepackage{babel}
\usepackage{amsfonts}
\usepackage{slashed}
\usepackage{enumerate}
\usepackage{soul}
\graphicspath{{./}}

\usepackage{babel}

\makeatother

\begin{document}
\global\long\def\sgn{\mathrm{sgn}}
\global\long\def\ket#1{|#1\rangle}
\global\long\def\bra#1{\langle#1|}
\global\long\def\sp#1#2{\langle#1|#2\rangle}
\global\long\def\abs#1{\left|#1\right|}

\title{Universal Quantum Noise in Adiabatic Pumping}

\author{Yaroslav Herasymenko}

\affiliation{Instituut-Lorentz for Theoretical Physics, Leiden University, Leiden,
NL-2333 CA, The Netherlands}

\author{Kyrylo Snizhko}

\affiliation{Department of Condensed Matter Physics, Weizmann Institute of Science,
Rehovot, 76100 Israel}

\author{Yuval Gefen}

\affiliation{Department of Condensed Matter Physics, Weizmann Institute of Science,
Rehovot, 76100 Israel}

\date{\today}
\begin{abstract}
We consider charge pumping in a system of parafermions, implemented
at fractional quantum Hall edges. Our pumping protocol leads to a
noisy behavior of the pumped current. As the adiabatic limit is approached,
not only does the noisy behavior persist but the counting statistics
of the pumped current becomes robust and universal. In particular,
the resulting Fano factor is given in terms of the system's topological
degeneracy and the pumped quasiparticle charge. Our results are also
applicable to the more conventional Majorana fermions.
\end{abstract}
\maketitle

Adiabatic quantum pumping, first introduced by Thouless \cite{Thouless1983},
is a powerful instrument in studying properties of quantum systems.
The underlying physics can be related to the system's Berry phase
\cite{Thouless1983}, disorder configurations \cite{Spivak1995},
scattering matrix and transport \cite{Brouwer1998}, critical points
\cite{Sela2006}, and topological properties \cite{Fu2006,Kraus2012,Keselman2013,Marra2015}.
In many cases \cite{Thouless1983,Sela2006,Fu2006,Kraus2012,Keselman2013,Marra2015},
adiabatic pumping is noiseless at zero temperature, as the same number
of quanta (of charge, spin, etc.) is pumped every cycle and the pumping
precision is increased (the noise vanishes) as the adiabatic limit
is approached. On the other hand, noisy adiabatic quantum pumps are
known and have been extensively studied \cite{Andreev2000,Avron2001,Makhlin2001,Moskalets2002,Moskalets2004,Riwar2013}.
The simplest (and a typical) example of such a noisy pump is two reservoirs
of electrons connected by a junction described by a scattering matrix.
As the phase of the reflection amplitude $r$ is varied from $0$
to $2\pi$, an electron is pumped with probability $|r|^{2}$ \cite{Andreev2000}.
The probabilistic nature of the adiabatic pumping process relies on
the degeneracy of scattering states. The pumped current and its noise
are sensitive to $|r|$, which in turn is highly sensitive to the
system parameters. In fact, in all such examples \cite{Andreev2000,Avron2001,Makhlin2001,Moskalets2002,Moskalets2004,Riwar2013},
the pumped current and its noise depend on the details of the pumping
cycle and/or of coupling the system to external leads.

In this Letter, we implement the concept of adiabatic pumping to a
setup of topological matter. We find that, when the adiabatic limit
is approached, not only is the pumped current noisy (a manifestation
of the degeneracy of the underlying Hilbert space), but it is also
universal: the current and its noise become largely independent of
the specific parameters used in the pumping cycle, and the related
Fano factor is directly related to the underlying topological structure;
cf.\,Eq.\,(\ref{eq:univ_current_noise}). Before going into technical
details, we now summarize the essence and the physical origin of our
findings. 

\begin{figure}
\noindent \begin{centering}
\includegraphics[width=1\columnwidth]{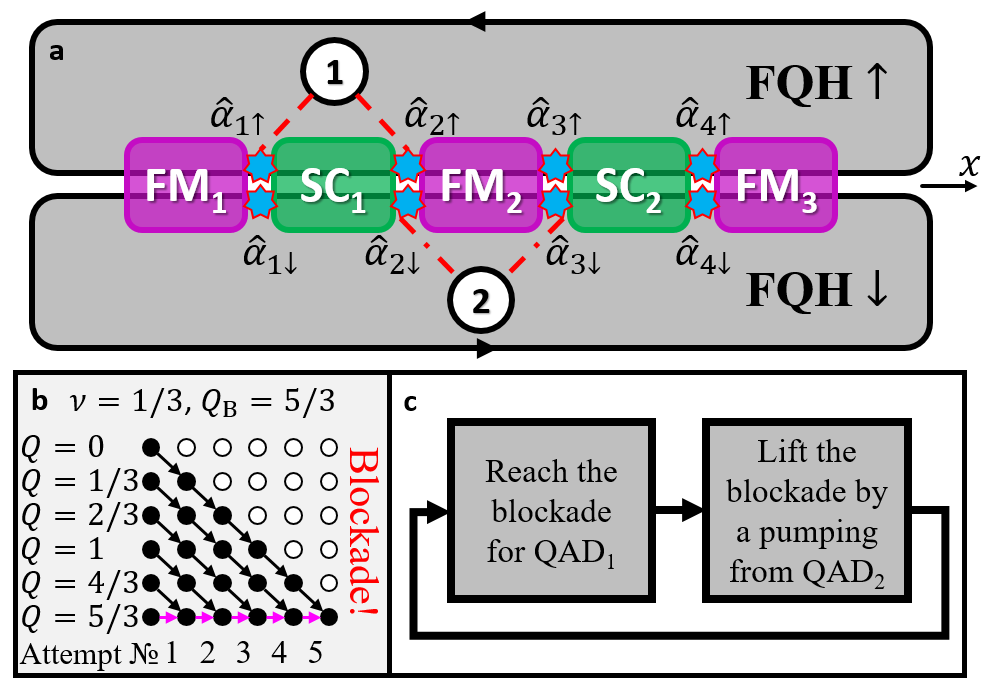}
\par\end{centering}
\caption{\label{fig:1}a\,— The system layout. In the regions proximitized
by ferromagnets (FM) and superconductors (SC), the FQH edges (of opposite
spin FQH puddles each of the same filling factor $\nu$) are gapped
out in two respective distinct ways. Each domain wall between a SC
and a FM region hosts PF zero mode operators (blue stars). The free
edges of spin-$\uparrow$ and spin-$\downarrow$ parts are glued together
by total reflection at the FMs. The bulk of the FQH puddles hosts
quantum anti-dots (QADs, denoted as 1 and 2)\,— regions depleted
by local gates. QADs behave as local enclaves that can support FQH
QPs. Tunnel couplings (red dashed and dot-dashed lines) between QADs
and parafermionic domain walls allow QPs to tunnel between them, influencing
the state of the PFs. All the proximitizing SCs (FMs) are implied
to be parts of a single bulk SC (FM), respectively. b\,— The mechanism
of $\mathrm{QAD}_{1}$ pumping blockade. Under repeated pumping attempts,
the system eventually reaches the state of $\mathrm{SC}_{1}$ domain
charge $Q=0$, in which pumping is blockaded. c\,— The elementary
cycle of the protocol producing universal pumping noise.}
\end{figure}

\emph{Qualitative overview of our protocol.—}The topological system
underlying our adiabatic pump is an array of parafermions (PFs), depicted
in Fig.\,\ref{fig:1}a. Consider an example of the system employing
fractional quantum Hall (FQH) puddles of filling factor $\nu=1/3$.
Each of the superconducting (SC) domains, $\mathrm{SC}_{i}$, is characterized
by the fractional component of its charge $Q_{i}/e=(0,1/3,2/3,...,5/3)$,
defined modulo $2e$ as charge quanta of $2e$ can be absorbed by
the proximitizing SC. Each of the two SC domains in Fig.\,\ref{fig:1}a
thus has $d=6$ states \footnote{Our protocol is also applicable to Majorana fermions that can be obtained
employing $\nu=1$ quantum Hall puddles or more conventional nanowires
\cite{Lutchyn2010,Oreg2010}. Then each SC domain/nanowire has $d=2$
states corresponding to $Q_{i}/e=0$ or $1$. Instead of fractional
quasiparticles, one would then pump electrons.}\nocite{Lutchyn2010,Oreg2010}. The system's topological nature renders
the states of different $Q_{i}$ degenerate, leading to $d^{2}$-degenerate
Hilbert space. Let us now consider a coherent source that is capable
of injecting FQH quasiparticles (QPs) of charge $e^{*}=e/3$ into
$\mathrm{SC}_{1}$. As the coherent source of QPs, we employ a quantum
antidot (QAD) \cite{Kivelson1989,Kivelson1990,Goldman1995,Goldman1997,Maasilta1997},
which is a depleted region in the FQH incompressible puddle that can
host fractional QPs. At low energies, this injection can take place
only at domain walls between $\mathrm{SC}_{1}$ and the neighboring
$\mathrm{FM}$ domains. As a result of such an injection, $Q\equiv Q_{1}$
would change $Q\rightarrow(Q+1/3)\thinspace\mathrm{\mod\thinspace2}$.
The two trajectories of injection (through the left or the right domain
wall) interfere with each other, implying that the probability of
a successful injection may be smaller than $1$ (and even tuned to
0). The latter, $P(Q)$, depends on the domain charge $Q$. $\mathrm{QAD}_{1}$
used for the injection of QPs into $\mathrm{SC}_{1}$ is denoted as
1 in Fig.\,\ref{fig:1}a.

It turns out that in the limit of adiabatic manipulation with the
QAD parameters, $P(Q)$ can be either $0$ when the interference is
fully destructive or $1$ otherwise {[}see the discussion after Eq.\,(\ref{eq:LZ_probability}){]}.
By tuning $P(Q=Q_{\mathrm{B}})=0$ for one of the system states $Q_{\mathrm{B}}$,
while $P(Q\neq Q_{\mathrm{B}})=1$, one blockades the repeated injection
of QPs as shown in Fig.\,\ref{fig:1}b: starting from any state,
the system eventually arrives in $Q=Q_{\mathrm{B}}$, stopping any
further injection of quasiparticles. We dub this phenomenon a \emph{topological
pumping blockade} \footnote{Cf.\,Refs.\,\cite{Flensberg2011,VanHeck2013,Kamenev2015} which
address the phenomenon of topological blockade, albeit not in the
context of pumping.}. \nocite{Flensberg2011,VanHeck2013, Kamenev2015}

We now employ an additional QAD ($\mathrm{QAD}_{2}$, denoted as 2
in Fig.\,\ref{fig:1}a) for lifting the blockade. A QP from $\mathrm{QAD}_{2}$
may be injected to either the second or the third domain wall. In
the former case it would change the $\mathrm{SC}_{1}$ charge $Q_{\mathrm{B}}\rightarrow(Q_{\mathrm{B}}+1/3)\thinspace\mathrm{\mod\thinspace2}$,
allowing for several more successful injections from $\mathrm{QAD}_{1}$,
while in the latter case the QP is injected to $\mathrm{SC}_{2}$,
leaving $Q$ unchanged. The probability of each outcome is governed
by the QP tunneling amplitude from $\mathrm{QAD}_{2}$ to the respective
domain wall. Consider a protocol whose elementary cycle consists of
$d-1$ QP injection attempts from $\mathrm{QAD}_{1}$ (sufficiently
many to reach the blockade irrespectively of the system initial state)
followed by disconnecting $\mathrm{QAD}_{1}$ from the array, then
a single injection from $\mathrm{QAD}_{2}$, and finally disconnecting
$\mathrm{QAD}_{2}$; cf.\,Fig.\,\ref{fig:1}c. Then in each cycle
the number of qps successfully injected from $\mathrm{QAD}_{1}$ is
determined by the value of $Q$ at the beginning of the cycle and
should therefore be either 0 or 5 with the corresponding probabilities.

A more careful consideration, however, shows that the mere connection
of $\mathrm{QAD}_{2}$ to the two domain walls simultaneously allows
for transfer of QPs between $\mathrm{SC}_{1}$ and $\mathrm{SC}_{2}$:
a QP can jump (through a virtual or a real process) from one domain
wall to the QAD and then to the other domain wall. As a result, any
state $Q$ at the beginning of the cycle is possible. For example,
if the QP from $\mathrm{QAD}_{2}$ is injected to $\mathrm{SC}_{1}$
and on top of that $k$ QPs are transferred from $\mathrm{SC}_{2}$
to $\mathrm{SC}_{1}$, then $Q_{\mathrm{B}}\rightarrow\left(Q_{\mathrm{B}}+(k+1)/3\right)\thinspace\mathrm{\mod\thinspace2}$.
Moreover, transfers of $k$ and $k+d$ QPs lead to the same value
of $Q$, and, therefore, these processes interfere. The interference
phases of these processes are sensitive to such parameters as the
strength of tunneling amplitudes between $\mathrm{QAD}_{2}$ and the
domain walls, the QAD potential, or the duration of the injection
process. In the adiabatic limit, a tiny cycle-to-cycle variation of
these parameters leads to a strong variation of the interference phases.
Therefore, averaged over many pumping cycles, the probability of starting
the cycle in any of the $d$ possible states $Q$ is the same and
is equal to $1/d$. The average current of charge pumped from $\mathrm{QAD}_{1}$
into the array, $I$, and its zero-frequency noise $S$, are then
given, respectively, by
\begin{gather}
I=I_{0}\frac{d-1}{2d},\quad S=\frac{d+1}{6}e^{*}I,\label{eq:univ_current_noise}
\end{gather}
where $I_{0}=e^{*}/\tau$ and $\tau$ is the duration of a single
injection attempt.

\emph{The model. Parafermions.—}Following Refs.\,\cite{Lindner2012,Clarke2013},
we consider a parafermion array realized on the boundary of two $\nu=1/(2p+1)$
FQH puddles, consisting of electrons of opposite spin; cf.\,Fig.\,\ref{fig:1}a.
The dynamics of the respective FQH edges is described by fields $\hat{\phi}_{s}(x)$,
$s=\pm1=\uparrow/\downarrow$, satisfying $[\hat{\phi}_{s}(x),\hat{\phi}_{s}(y)]=i\pi s\sgn(x-y)$
and $[\hat{\phi}_{\uparrow}(x),\hat{\phi}_{\downarrow}(y)]=i\pi$
\cite{Clarke2013}. The edges support domains that are gapped by proximity
coupling to a superconductor (SC) or a ferromagnet (FM); $H=H_{\mathrm{edge}}+H_{\mathrm{SC}}+H_{\mathrm{FM}}$,
where $H_{\mathrm{edge}}=(v/4\pi)\int_{0}^{L}dx\left[(\partial_{x}\hat{\phi}_{\uparrow})^{2}+(\partial_{x}\hat{\phi}_{\downarrow})^{2}\right]$
with edge velocity $v$,

\begin{eqnarray}
H_{\mathrm{SC}} & = & -\frac{\Delta}{a}\sum_{j=1}^{N}\int_{\mathrm{SC}_{j}}dx\cos\left(\frac{\hat{\phi}_{\uparrow}(x)+\hat{\phi}_{\downarrow}(x)}{\sqrt{\nu}}\right),\label{eq:H_SC}\\
H_{\mathrm{FM}} & = & -\frac{\mathcal{M}}{a}\sum_{j=1}^{N+1}\int_{\mathrm{FM}_{j}}dx\cos\left(\frac{\hat{\phi}_{\uparrow}(x)-\hat{\phi}_{\downarrow}(x)}{\sqrt{\nu}}\right),\label{eq:H_FM}
\end{eqnarray}

\noindent with $\Delta$ (respectively, $\mathcal{M}$) being the
absolute value of the induced amplitude for SC pairing (for tunneling
between edge segments proximitized by FMs), short-distance cutoff
$a$, and $N=2$ is the number of SC domains. All the proximitizing
SCs (FMs) are implied to be parts of a single bulk SC (FM), respectively.
The bulk SC is assumed to be grounded. For $\Delta a/v$, $\mathcal{M}a/v>\sqrt{2\nu-\ln2\nu-1}/(2\sqrt{2}\pi\nu^{2})$
when $\nu\leq1/3$ \footnote{The expressions follow from the analysis of renormalization group
(RG) equations for a single infinite domain. The Hamiltonian for a
single domain is essentially that of the sine-Gordon model, and the
RG flow is that of the Berezinskii-Kosterlitz-Thouless transition
\cite[section 8.6]{Altland2010}.}\nocite{Altland2010} and for any nonzero values of $\Delta a/v$
and $\mathcal{M}a/v$ when $\nu=1$, each domain has a gap for QP
excitations. At low energies, each domain can be described by a single
integer-valued operator\,\cite{Lindner2012,Clarke2013}
\begin{equation}
\left.\frac{\hat{\phi}_{\uparrow}(x)\mp\hat{\phi}_{\downarrow}(x)}{2\pi\sqrt{\nu}}\right|_{x\in\mathrm{FM}_{j}/\mathrm{SC}_{j}}=\left\{ \begin{array}{c}
\hat{m}_{j},\\
\hat{n}_{j}.
\end{array}\right.
\end{equation}
The only nontrivial commutation relation is $[\hat{m}_{j},\hat{n}_{l}]=i/(\pi\nu)$
for $j>l$, while $[\hat{m}_{j},\hat{n}_{l}]=0$ for $j\le l$. Being
integer-valued noncommuting operators, they are defined modulo $d=2/\nu$,
i.e., $\hat{m}_{j}\thinspace(\hat{n}_{j})\sim\hat{m}_{j}\thinspace(\hat{n}_{j})+d$.
The fractional component of the $j^{\mathrm{th}}$ SC domain's charge
$\hat{Q}_{j}$ is given by $\hat{Q}_{j}\thinspace\mathrm{mod}\thinspace2e=e^{*}(\hat{m}_{j+1}-\hat{m}_{j})\thinspace\mathrm{mod}\thinspace2e=\nu\left[(\hat{m}_{j+1}-\hat{m}_{j})\thinspace\mathrm{mod}\thinspace d\right]$,
where $e^{*}=\nu e$ and $e$ are, respectively, the charge of the
fractional QP and the electron charge, and we put $e=1$. The parafermion
array Hilbert space may be spanned by states $|m_{1},Q,m_{3}\rangle$,
where $m_{j}$ is the eigenvalue of $\hat{m}_{j}$ and $Q$ is the
eigenvalue of $(\hat{Q}_{1}\thinspace\mathrm{mod}\thinspace2e)$.
Alternatively, one can use the basis of $|m_{1},S,m_{3}\rangle$ with
$S$ being the eigenvalue of $\nu\left[(\hat{n}_{1}-\hat{n}_{2})\thinspace\mathrm{mod}\thinspace d\right]$.
The possible values for both $Q$ and $S$ are $0,\nu,...,(d-1)\nu\equiv2-\nu$
\footnote{\noindent  For the sake of brevity, in the formulas below we allow
values of $Q$ and $S$ beyond the interval $\left[0;2-\nu\right]$,
implying that those are shifted to this interval by taking them $\mathrm{mod}\thinspace2$.}. These two bases are related as
\begin{equation}
\ket{m_{1},S,m_{3}}=\frac{1}{\sqrt{d}}\sum_{Q=0}^{(d-1)\nu}e^{i\pi dQS/2}\ket{m_{1},Q,m_{3}}.\label{eq:bases_relation}
\end{equation}

Our protocols involve tunneling fractional QPs into the parafermion
array. At low energies such tunneling may take place only at the interfaces
between different domains. The low-energy projection of the QP operators
is given by (cf.\,Refs.\,\cite{Lindner2012,Clarke2013})
\begin{equation}
\hat{\alpha}_{js}=\left\{ \begin{array}{cc}
e^{i\pi\nu(\hat{n}_{l}+s\hat{m}_{l})}, & j=2l-1,\\[0.1cm]
e^{i\pi\nu(\hat{n}_{l}+s\hat{m}_{l+1})}, & j=2l,
\end{array}\right.,\label{eq:PF_explicit_form}
\end{equation}

\noindent where $j$ is the domain wall number and $s=\pm1=\uparrow/\downarrow$
is the spin of the edge into which the QP tunnels. For $\nu=1$, $\hat{\alpha}_{js}$
become Majorana fermions.

In addition to the parafermion-hosting domain walls, \emph{quantum
antidots} are the second main ingredient of our model. We consider
small QADs in the Coulomb blockade regime. Such a QAD can be modeled
as a system of two levels, $|q\rangle$ and $|q+\nu\rangle$, corresponding
to the QAD hosting charge $q$ or $q+\nu$ respectively. The QP operator
on the QAD and the QAD Hamiltonian assume then the forms
\begin{gather}
\hat{\psi}_{\mathrm{QAD}}=\begin{pmatrix}0 & 0\\
1 & 0
\end{pmatrix},\\
H_{\mathrm{QAD}}=\nu V_{\mathrm{QAD}}\left(\hat{\psi}_{\mathrm{QAD}}^{\dagger}\hat{\psi}_{\mathrm{QAD}}^{\thinspace}-\frac{1}{2}\right)=\frac{V_{\mathrm{QAD}}}{d}\begin{pmatrix}1 & 0\\
0 & -1
\end{pmatrix},\label{eq:H_QAD}
\end{gather}

\noindent where $V_{\mathrm{QAD}}$ is an electrostatic gate potential.
One can consider several QADs, each described by such a two-level
Hamiltonian \footnote{In principle, one has to introduce Klein factors to ensure appropriate
permutation relations between the QP operators of different QADs and
also between the QP operators and the PFs. However, it turns out that
the Klein factors do not influence the physical observables in the
present analysis. Indeed, they multiply the QAD QP operator by a phase
that depends on the total charge of the PF system and on the occupation
of the other QADs. However, these phase factors do not influence the
observables in the proposed protocol.}.

The Hamiltonian describing tunneling of QPs between a QAD and the
PF system is
\begin{equation}
H_{\mathrm{tun}}=\sum_{j}\eta_{js}\hat{\psi}_{\mathrm{QAD},s}^{\thinspace}\hat{\alpha}_{js}^{\dagger}+\mathrm{H.c.}\label{eq:H_tun}
\end{equation}

\noindent Here $\eta_{js}$ is the tunneling amplitude to the $j^{\mathrm{th}}$
domain wall, and $\hat{\alpha}_{js}$ is the PF operator in this domain
wall. Fractional QPs can tunnel only through a FQH bulk but not through
a vacuum. The QAD is embedded in the FQH puddle of spin $s$ and is
therefore coupled only to the PFs of the same spin; this is indicated
by index $s$ of the QAD operator.

\emph{Injection of a QP from $\mathrm{QAD}_{1}$}.—In Fig.\,\ref{fig:1}a\emph{,
$\mathrm{QAD}_{1}$} is connected to parafermions $\hat{\alpha}_{1\uparrow}$
and $\hat{\alpha}_{2\uparrow}$. The tunneling Hamiltonian (\ref{eq:H_tun})
then allows for transitions only between states $|q+\nu\rangle_{\mathrm{QAD}_{1}}\ket{m_{1},Q,m_{3}}\equiv|1\rangle$
and $|q\rangle_{\mathrm{QAD}_{1}}\ket{m_{1},Q+\nu,m_{3}+1}\equiv|0\rangle$.
The problem of QP tunneling can therefore be mapped onto a set of
$2\times2$ problems each described by the Hamiltonian

\begin{gather}
H_{\mathrm{LZ}}(t)=\begin{pmatrix}\frac{1}{d}V_{\mathrm{QAD}}(t) & \eta_{Q}^{*}\\
\eta_{Q} & -\frac{1}{d}V_{\mathrm{QAD}}(t)
\end{pmatrix},\label{eq:LZ_Hamiltonian}\\
\eta_{Q}=e^{-i\pi\nu m_{1}}\left(\eta_{1\uparrow}+\eta_{2\uparrow}e^{-i\pi\left(Q+\frac{\nu}{2}\right)}\right).\label{eq:eta_Q}
\end{gather}
For this Hamiltonian, consider the Landau-Zener problem \cite{LZ_refs,Landau1977}:
$V_{\mathrm{QAD}}(t)=\nu^{-1}\lambda t$ with $\lambda>0$; at $t=-T$
the effective two-level system is prepared in the lower-energy state
$|\psi(-T)\rangle=|1\rangle$ ($|1\rangle$ and $|0\rangle$ are the
\emph{diabatic} states of the QAD–PF system). Then at $t=+T$ it will
generally be in a superposition of the two diabatic states. When $T\rightarrow+\infty$,
the probability of staying in state $|1\rangle$ (i.e., not injecting
the QP) is

\begin{equation}
P_{\mathrm{LZ}}=\exp\left(-2\pi\gamma\right),\label{eq:LZ_probability}
\end{equation}

\noindent where $\gamma=|\eta_{Q}|^{2}/\lambda$. Unless $\eta_{Q}=0$,
the probability $P(Q)=1-P_{\mathrm{LZ}}$ of switching from $\ket 1$
to $\ket 0$, i.e., of injecting a QP to $\mathrm{SC}_{1}$ domain,
is exponentially close to 1 in the adiabatic limit ($\lambda\rightarrow0$,
the limiting QAD potential $V_{0}=\nu^{-1}\lambda T=\mathrm{const}\gg\max_{Q}|\eta_{Q}|$).
By fine-tuning $\eta_{1\uparrow}/\eta_{2\uparrow}=-e^{-i\pi\left(Q_{\mathrm{B}}+\frac{\nu}{2}\right)}$
with a certain $Q_{\mathrm{B}}=0,\nu,...,2-\nu$, one achieves $P(Q_{\mathrm{B}})=0$.
If the fine-tuning is imperfect, the precision of $P(Q_{\mathrm{B}})=0$
is determined by how well $\eta_{Q_{\mathrm{B}}}$ is tuned to zero:
$|\eta_{Q_{\mathrm{B}}}|\leq\sqrt{C\lambda}$ implies $P(Q_{\mathrm{B}})\leq1-e^{-2\pi C}\leq2\pi C$.
Summing up, in the adiabatic limit an injection attempt is either
successful with unit probability or has zero probability of success
depending on the system state $Q$ and the tunneling amplitudes' ratio
$\eta_{1\uparrow}/\eta_{2\uparrow}$. Below, we employ \emph{$\mathrm{QAD}_{1}$}
with the above fine-tuned tunneling amplitudes. A successful injection
implies $\ket{m_{1},Q,m_{3}}\rightarrow e^{i\theta_{Q}}\ket{m_{1},Q+\nu,m_{3}+1}$
with phases $\theta_{Q}$ that are unimportant to us, while an unsuccessful
one implies $\ket{m_{1},Q_{\mathrm{B}},m_{3}}\rightarrow\ket{m_{1},Q_{\mathrm{B}},m_{3}}$.

The origin of the \emph{topological pumping blockade} {[}Fig.\,\ref{fig:1}b{]}
now becomes clear. Define a pumping (injection) attempt as preparing
\emph{$\mathrm{QAD}_{1}$} in the state $|q+\nu\rangle_{\mathrm{QAD}_{1}}$,
connecting \emph{$\mathrm{QAD}_{1}$} to parafermions, adiabatically
sweeping $V_{\mathrm{QAD}}$ from $-V_{0}$ to $V_{0}$, and disconnecting
the QAD from the array. Prepare the array in a generic superposition
of $Q$-states. A single injection attempt transforms the initial
state of the QAD and parafermions: 
\begin{multline}
|q+\nu\rangle_{\mathrm{QAD}_{1}}\sum_{Q=0}^{2-\nu}A_{Q}\ket{m_{1},Q,m_{3}}\rightarrow\\
|q+\nu\rangle_{\mathrm{QAD}_{1}}A_{0}\ket{m_{1},2-\nu,m_{3}}\\
+|q\rangle_{\mathrm{QAD}_{1}}\sum_{Q=\nu}^{2-\nu}A_{Q-\nu}e^{i\theta_{Q-\nu}}\ket{m_{1},Q,m_{3}+1},\label{eq:pumping_attempt_state_evolution}
\end{multline}
where we assumed without the loss of generality that $Q_{\mathrm{B}}=2-\nu$.
The injection attempt will be unsuccessful (projecting the state to
$|Q=Q_{\mathrm{B}}\rangle$) with probability $|A_{0}|^{2}$, while
with probability $1-|A_{0}|^{2}$ the pumping attempt will be successful,
resulting in the $Q$-state being a superposition of $\ket{m_{1},Q,m_{3}+1}$,
$Q=\nu,...,2-\nu$. After $k-1$ such attempts, the array will be
either in the state with $Q=Q_{\mathrm{B}}$ or in a superposition
of $Q$ between $(k-1)\nu$ and $2-\nu\equiv(d-1)\nu$. Following
$d-1$ pumping attempts, the array state will definitely have $Q=Q_{\mathrm{B}}$,
and further pumping will be blockaded {[}cf.\,Fig.\,\ref{fig:1}b{]}.

\begin{figure}
\noindent \begin{centering}
\includegraphics[width=1\columnwidth]{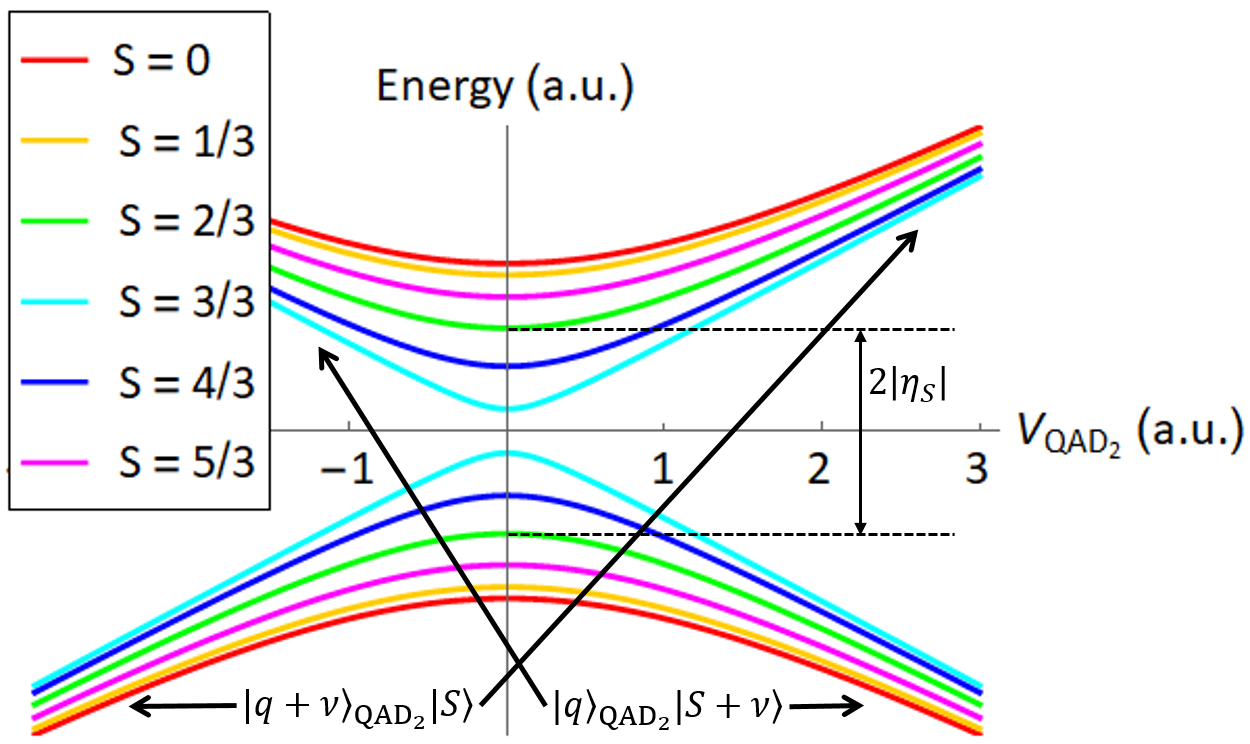}
\par\end{centering}
\caption{\label{fig:2}Energy of adiabatic states when injecting a quasiparticle
from $\mathrm{QAD}_{2}$. The states of different $S$ have different
energies and hence accumulate different dynamical phase during the
process. The sensitivity of the dynamical phase to the process parameters
is the origin of universal noise in our protocol.}
\end{figure}

Consider now in detail the process of \emph{injecting of a QP from
$\mathrm{QAD}_{2}$}. $\mathrm{QAD}_{2}$ is connected to parafermions
$\hat{\alpha}_{2\downarrow}$ and $\hat{\alpha}_{3\downarrow}$, rendering
$\ket{m_{1},S,m_{3}}$ a convenient basis to work with. Indeed, the
tunneling Hamiltonian (\ref{eq:H_tun}) allows for transitions only
between states $|q+\nu\rangle_{\mathrm{QAD}_{2}}\ket{m_{1},S,m_{3}}\equiv|1\rangle$
and $|q\rangle_{\mathrm{QAD}_{2}}\ket{m_{1},S+\nu,m_{3}+1}\equiv|0\rangle$.
In this basis, tunneling from $\mathrm{QAD}_{2}$ is described by
the same Hamiltonian as in (\ref{eq:LZ_Hamiltonian}) except $\eta_{Q}$
should be replaced with
\begin{equation}
\eta_{S}=e^{i\pi\nu m_{1}}\left(\eta_{2\downarrow}e^{-i\pi\left(S+\frac{\nu}{2}\right)}+\eta_{3\downarrow}\right).
\end{equation}
The physics of injecting a QP from $\mathrm{QAD}_{2}$ is therefore
similar to that of injection from \emph{$\mathrm{QAD}_{1}$}. However,
we employ $\mathrm{QAD}_{2}$ only in the non-blockaded regime. In
other words, $\eta_{S}\neq0$ for all $S$. Therefore, in the adiabatic
limit the injection is always successful, implying $\ket{m_{1},S,m_{3}}\rightarrow e^{i\theta_{S}}\ket{m_{1},S+\nu,m_{3}+1}$
with phases
\begin{equation}
\theta_{S}=\frac{(\nu V_{0})^{2}}{2\lambda}-\pi-i\ln\frac{\eta_{S}}{|\eta_{S}|}+\frac{|\eta_{S}|^{2}}{\lambda}\left(1+\ln\frac{(\nu V_{0})^{2}}{|\eta_{S}|^{2}}\right).
\end{equation}
These phases are of utmost importance for our protocol. The terms
proportional to $\lambda^{-1}$ can be understood as dynamical phases
$-\int_{-T}^{T}E_{S}(t)dt$ associated with the adiabatic states of
the process having energies $E_{S}(t)=-\sqrt{\abs{\eta_{S}}^{2}+\left(V_{\mathrm{QAD}}(t)/d\right)^{2}}$;
cf.\,Fig.\,\ref{fig:2}. In the adiabatic limit $\lambda\rightarrow0$,
these terms tend to infinity. As a result, the phase is highly sensitive
even to the tiniest variations of the parameters involved. For a example,
a small change $\delta V_{0}\ll V_{0}$ of the limiting QAD potential
$V_{0}$ modifies the phase by
\begin{equation}
\delta\theta_{S}=\frac{(\nu V_{0})^{2}}{\lambda}\frac{\delta V_{0}}{V_{0}}+2\frac{|\eta_{S}|^{2}}{\lambda}\frac{\delta V_{0}}{V_{0}},\label{eq:phase_fluct}
\end{equation}
which diverges in the adiabatic limit.

We are now in a position to discuss the\emph{ pumping protocol} whose
cycle is schematically shown in Fig.\,\ref{fig:1}c. After the sequence
of injection attempts from $\mathrm{QAD}_{1}$, the system evolves
into a state with $Q=Q_{\mathrm{B}}$, say, $\ket{m_{1},Q_{B},m_{3}}$.
The injection of a QP from $\mathrm{QAD}_{2}$ evolves this state
to
\begin{multline}
\sum_{S=0}^{2-\nu}e^{i\theta_{S}}\ket{m_{1},S+\nu,m_{3}+1}\bra{m_{1},S,m_{3}}m_{1},Q_{\mathrm{B}},m_{3}\rangle\\
=\sum_{Q}A_{Q}\ket{m_{1},Q,m_{3}+1},
\end{multline}
\begin{equation}
A_{Q}=\frac{1}{d}\sum_{S=0}^{2-\nu}e^{i\pi d(Q-Q_{\mathrm{B}})S/2+i\pi Q+i\theta_{S}}.
\end{equation}
Therefore, the probability of pumping $r$ QPs from $\mathrm{QAD}_{1}$
in the next pumping cycle is given by $\abs{A_{Q=Q_{\mathrm{B}}-r\nu}}^{2}$.

Assume that in each pumping cycle the limiting $\mathrm{QAD}_{2}$
potential $V_{0}$ is slightly different. The phases $\theta_{S}$
exhibit then cycle-to-cycle fluctuations; we are interested in the
probabilities $\abs{A_{Q=Q_{\mathrm{B}}-r\nu}}^{2}$ averaged over
these fluctuations:
\begin{equation}
\langle\abs{A_{Q}}^{2}\rangle_{\delta V_{0}}=\frac{1}{d^{2}}\sum_{S,S'=0}^{2-\nu}e^{i\pi d(Q-Q_{\mathrm{B}})(S-S')/2}\langle e^{i(\theta_{S}-\theta_{S'})}\rangle_{\delta V_{0}}.
\end{equation}
Note that 
\begin{equation}
\delta\theta_{S}-\delta\theta_{S'}=2\frac{|\eta_{S}|^{2}-|\eta_{S'}|^{2}}{\lambda}\frac{\delta V_{0}}{V_{0}}
\end{equation}
diverges in the adiabatic limit for arbitrarily small fluctuations
$\delta V_{0}$, provided that $|\eta_{S}|\neq|\eta_{S'}|$; the latter
is generically true. Hence, $\langle e^{i(\theta_{S}-\theta_{S'})}\rangle_{\delta V_{0}}=0$
for $S\neq S'$ and $\langle\abs{A_{Q}}^{2}\rangle_{\delta V_{0}}=1/d$.
Therefore, the number of QPs pumped from $\mathrm{QAD}_{1}$ in each
cycle has a universal probability distribution, leading to a universal
counting statistics of the pumping current. In particular, the average
current and the zero-frequency noise are given by Eq.\,(\ref{eq:univ_current_noise}).

\emph{Discussion.}—The topological nature of our parafermion system
gives rise to a degenerate set of ``scattering states''. The latter
render charge pumping in the adiabatic limit noisy. In sharp contrast
to earlier studies of noisy pumping, here the average current as well
as the noise (and, in fact, the entire counting statistics) are found
to be topology-related universal. Specifically, the Fano factor $(d+1)e^{*}/6$
is directly related to the topological degeneracy $d$ of the parafermionic
space. In analogy with the quantum Hall effect, where static disorder
is needed to provide robustness to the quantized Hall conductance,
here we require (minute) time-dependent (cycle-to-cycle) variations
of the pumping parameters used for $\mathrm{QAD}_{2}$. Majorana zero
modes are a special case of our protocol ($d=2$). In that case, the
system does not support fractional quasiparticles, and one pumps electrons
(rather than fractionally charged anyons) into the array of topological
modes; therefore, conventional quantum dots (rather than quantum antidots
embedded in FQH puddles) can be employed. For realizing the Majorana
array, one can use the boundary between two $\nu=1$ quantum Hall
puddles or, alternatively, a set of Majorana wires. The Fano factor
will then be $1/2$.

\begin{acknowledgments}

\emph{Acknowledgments}. K.\,S. thanks A.\,Haim for discussions.
Y.\,H. thanks the Kupcinet-Getz program at Weizmann Institute of
Science during participation in which he joined this project. We acknowledge
funding by the Deutsche Forschungsgemeinschaft (Bonn) within the network
CRC TR 183 (Project No. C01) and Grant No.~RO 2247/8-1, by the ISF,
and the Italia-Israel project QUANTRA. Y.\,G. acknowledges funding
by the IMOS Israel-Russia program. This text was prepared with the
help of LyX software \cite{LyX}.

Y.\,H. and K.\,S. have made equal contributions. 

\end{acknowledgments}

\bibliographystyle{apsrev4-1}
\bibliography{C:/Users/ksnizhko.CMD-YGKYRYLO1/Documents/library,Suppl}

\end{document}